\documentclass[aps,prl,twocolumn,10pt,superscriptaddress]{revtex4-1}

\usepackage{amsmath, amssymb}
\usepackage[colorlinks=true,urlcolor=blue,citecolor=blue,linkcolor=blue]{hyperref}
\usepackage{graphicx}

\begin{document}

\title{Detector-Independent Verification of Quantum Light}

\author{J.~Sperling}\email{jan.sperling@physics.ox.ac.uk}
\affiliation{Clarendon Laboratory, University of Oxford, Parks Road, Oxford OX1 3PU, United Kingdom}

\author{W.~R.~Clements}
\affiliation{Clarendon Laboratory, University of Oxford, Parks Road, Oxford OX1 3PU, United Kingdom}

\author{A.~Eckstein}
\affiliation{Clarendon Laboratory, University of Oxford, Parks Road, Oxford OX1 3PU, United Kingdom}

\author{M.~Moore}
\affiliation{Clarendon Laboratory, University of Oxford, Parks Road, Oxford OX1 3PU, United Kingdom}

\author{J.~J.~Renema}
\affiliation{Clarendon Laboratory, University of Oxford, Parks Road, Oxford OX1 3PU, United Kingdom}

\author{W.~S.~Kolthammer}
\affiliation{Clarendon Laboratory, University of Oxford, Parks Road, Oxford OX1 3PU, United Kingdom}

\author{S.~W.~Nam}
\affiliation{National Institute of Standards and Technology, 325 Broadway, Boulder, Colorado 80305, USA}

\author{A.~Lita}
\affiliation{National Institute of Standards and Technology, 325 Broadway, Boulder, Colorado 80305, USA}

\author{T.~Gerrits}
\affiliation{National Institute of Standards and Technology, 325 Broadway, Boulder, Colorado 80305, USA}

\author{W.~Vogel}
\affiliation{Institut f\"ur Physik, Universit\"at Rostock, Albert-Einstein-Stra\ss{}e 23, D-18059 Rostock, Germany}

\author{G.~S.~Agarwal}
\affiliation{Texas A\&M University, College Station, Texas 77845, USA}

\author{I.~A.~Walmsley}
\affiliation{Clarendon Laboratory, University of Oxford, Parks Road, Oxford OX1 3PU, United Kingdom}

\date{\today}

\begin{abstract}
	We introduce a method for the verification of nonclassical light which is independent of the complex interaction between the generated light and the material of the detectors.
	This is accomplished by means of a multiplexing arrangement.
	Its theoretical description yields that the coincidence statistics of this measurement layout is a mixture of multinomial distributions for any classical light field and any type of detector.
	This allows us to formulate bounds on the statistical properties of classical states.
	We apply our directly accessible method to heralded multiphoton states which are detected with a single multiplexing step only and two detectors, which are in our work superconducting transition-edge sensors.
	The nonclassicality of the generated light is verified and characterized through the violation of the classical bounds without the need for characterizing the used detectors.
\end{abstract}

\maketitle

\paragraph*{Introduction.---}
	The generation and verification of nonclassical light is one of the main challenges for realizing optical quantum communication and computation \cite{KLM01,KMNRDM07,GT07,S09}.
	Therefore, robust and easily applicable methods are required to detect quantum features for real-world applications; see, e.g., \cite{TSDZDW11,BFM15}.

	The complexity of producing reliable sensors stems from the problem that new detectors need to be characterized.
	For this task, various techniques, such as detector tomography \cite{LS99,AMP04,LKKFSL08,LFCPSREPW09,ZDCJEPW12,BCDGMMPPP12}, have been developed.
	However, such a calibration requires many resources, for example, computational or numerical analysis, reference measurements, etc.
	From such complex procedures, the interaction between quantum light and the bulk material of the detector can be inferred and quantum features can be uncovered.
	Nevertheless, the verification of nonclassicality also depends on the bare existence of criteria that are applicable to this measurement.
	Here, we prove that detectors with a general response to incident light can be employed in an optical detection scheme, which is well characterized, to identify nonclassical radiation fields based on simple nonclassicality conditions.

	The concept of device independence has recently gained a lot of attention because it allows one to employ even untrusted devices; see, e.g., \cite{KW16}.
	For instance, device-independent entanglement witnesses can be used without relying on properties of the measurement system \cite{BRLG13,ZYM16}.
	It has been further studied to perform communication and computation tasks \cite{LKMBTZ14,GKW15}.
	Detector independence has been also applied to state estimation and quantum metrology \cite{CKS14,AGSB16} to gain knowledge about a physical system which might be too complex for a full characterization.

	In parallel, remarkable progress has been made in the field of well-characterized photon-number-resolving (PNR) detectors \cite{S07,H09}.
	A charge-coupled-device camera is one example of a system that can record many photons at a time.
	However, it also suffers inherent readout noise.
	Still, the correlation between different pixels can be used to infer quantum correlated light \cite{BDFL08,MMDL12,CTFLMA16}.
	Another example of a PNR device is a superconducting transition-edge sensor (TES) \cite{LMN08,BCDGLMPRTP12,RFZMGDFE12}.
	This detector requires a cryogenic environment, and its operation is based on superconductivity.
	Hence, a model for this detector would require the quantum mechanical treatment of a solid-state bulk material which interacts with a quantized radiation field in the frame of low-temperature physics.

	Along with the development of PNR detectors, multiplexing layouts define another approach to realize photon-number resolution \cite{PTKJ96,KB01,ASSBW03,FJPF03,CDSM07,SPDBCM07}.
	The main idea is that an incident light field, which consists of many photons, is split into a number of spatial or temporal modes, which consist of a few photons only.
	These resulting beams are measured with single-photon detectors which do not have any photon-number-resolution capacity.
	They can only discriminate between the presence (``click'') and absence of absorbed photons.
	Hence, the multiplexing is used to get some insight into the photon statistics despite the limited capacity of the individual detectors.
	With its resulting binomial click-counting statistics, one can verify nonclassical properties of correlated light fields \cite{SVA12,BDJDBW13,SBVHBAS15,HSPGHNVS16,SBDBJDVW16}.
	Recently, a multiplexing layout has been used in combination with TESs to characterize quantum light with a mean photon number of 50 and a maximum number of 80 photons for each of the two correlated modes \cite{HBLNGS15}.

	In this Letter, we formulate a method to verify nonclassical light with arbitrary detectors.
	This technique is based on a well-defined multiplexing scheme and individual detectors which can discriminate different measurement outcomes.
	The resulting correlation measurement is always described as a mixture of multinomial distributions in classical optics.
	Based on this finding, we formulate nonclassicality conditions in terms of covariances to directly certify nonclassical light.
	Nonclassical light is defined in this work as a radiation field which cannot be described as a statistical mixture of coherent light \cite{TG86,M86}.
	We demonstrate our approach by producing heralded photon-number states from a parametric down-conversion (PDC) source.
	Already a single multiplexing step is sufficient to verify the nonclassicality of such states without the need to characterize the used TESs.
	In addition to our method presented here, a complementary study is provided in Ref. \cite{TheArticle}.
	There we use a quantum-optical framework to perform additional analysis of the measurement layout under study.

\begin{figure}[ht]
	\includegraphics[width=6cm]{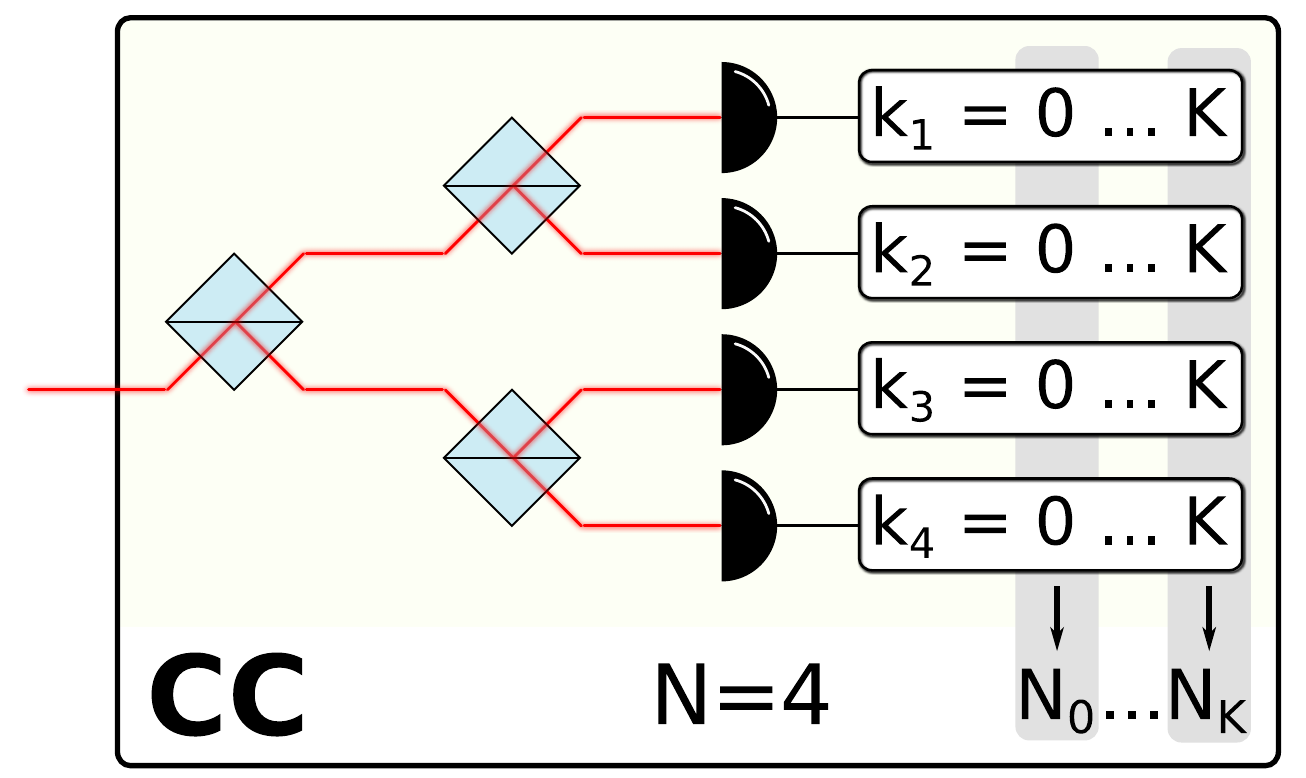}
	\caption{(Color online)
		Multiplexed click-counting (CC) layout consisting of $N=4$ individual detectors.
		Incident light is split into $N$ beams with similar intensities.
		Each of the $N$ detectors returns a measurement outcome $k_n$.
		The number of detectors $N_k$ with the same outcome $0\leq k\leq K$ is recorded.
	}\label{fig:theory}
\end{figure}

\paragraph*{Theory.---}
	The detection scenario is shown in Fig. \ref{fig:theory}.
	Its robustness to the detector response is achieved by the multiplexing layout whose optical elements, e.g., beam splitters, are much simpler and better characterized than the detectors.
	Our only broad requirement is that the measured statistics of the detectors are relatively similar to each other.
	Here we are not using multiplexing to improve the photon-number detection (see, e.g., Ref. \cite{HBLNGS15}).
	Rather, we employ this scheme to get nonclassicality criteria that are independent of the properties of the individual detectors.

	First, we consider a single coherent, classical light field.
	The detector can resolve arbitrary outcomes $k=0,\ldots,K$---or, equivalently, $K+1$ bins \cite{CommentBin}---which have a probability $p_k$.
	If the light is split by $50{/}50$ beam splitters as depicted in Fig. \ref{fig:theory} and measured with $N$ individual and identical detectors, we get the probability $p_{k_1}\cdots p_{k_N}$ to measure $k_1$ with the first detector, $k_2$ with the second detector, etc.
	Now, $N_k$ is defined as the number of individual detectors which measure the same outcome $k$.
	This means we have $N_0$ times the outcome $0$ together with $N_1$ times the outcome $1$, etc., from the $N$ detectors, $N=N_0+\cdots+N_K$.
	For example, $k_1=K$ and $k_2=k_3=k_4=0$ yields $N_K=1$ and $N_0=3$ for $N=4$ detectors ($N_k=0$ for all $0<k<K$).
	The probability to get any given combination of outcomes, $N_0,\ldots,N_K$, from the probabilities $p_{k_1}\cdots p_{k_N}$ is known to follow a multinomial distribution \cite{FEHP11},
	\begin{align}\label{eq:MultinomialDistribution}
		c(N_0,\ldots,N_K)=\frac{N!}{N_0!\cdots N_K!}p_0^{N_0}\cdots p_K^{N_K}.
	\end{align}
	To ensure a general applicability, we counter any deviation from the $50{/}50$ splitting and differences of the individual detectors by determining a corresponding systematic error (in our experiment in the order of $1\%$), see the Supplemental Material \cite{SupplementalMaterial} for the error analysis.

	For a different intensity, the probabilities $p_k$ of the individual outcomes $k$ might change.
	Hence, we consider in the second step a statistical mixture of arbitrary intensities.
	This generalizes the distribution in Eq. \eqref{eq:MultinomialDistribution} by averaging over a classical probability distribution $P$,
	\begin{align}\label{eq:AveragedMultinomialDistribution}
	\begin{aligned}
		&c(N_0,\ldots,N_K)=\left\langle\frac{N!}{N_0!\cdots N_K!}p_0^{N_0}\cdots p_K^{N_K}\right\rangle
		\\=&
		\int dP(p_0,\ldots,p_K)\frac{N!}{N_0!\cdots N_K!}p_0^{N_0}\cdots p_K^{N_K}.
	\end{aligned}
	\end{align}
	Because any light field in classical optics can be considered as an ensemble of coherent fields \cite{TG86,M86}, the measured statistics of the setup in Fig. \ref{fig:theory} follows a mixture of multinomial distributions \eqref{eq:AveragedMultinomialDistribution}.
	This is not necessarily true for nonclassical light as we will demonstrate.
	The distribution \eqref{eq:AveragedMultinomialDistribution} applies to arbitrary detectors and includes the case of on-off detectors ($K=1$), which yields a binomial distribution \cite{SVA12a}.
	Also, we determine the number of outcomes, $K+1$, directly from our data.

	Let us now formulate a criterion that allows for the identification of quantum correlations.
	The mean values of multinomial statistics obey $\overline{N_k}=Np_k$ \cite{FEHP11}.
	Averaging over $P$ yields
	\begin{align}\label{eq:firstorderclickmoments}
		\overline{N_k}=N\left\langle p_k\right\rangle.
	\end{align}
	In the same way, we get for the second-order moments, $\overline{N_kN_{k'}}=N(N-1)p_kp_{k'}+\delta_{k,k'}Np_k$ \cite{FEHP11} with $\delta_{k,k'}=1$ for $k=k'$ and $\delta_{k,k'}=0$ otherwise, an averaged expression
	\begin{align}\label{eq:secondorderclickmoments}
		\overline{N_kN_{k'}}=&N(N-1)\left\langle p_kp_{k'}\right\rangle+\delta_{k,k'}N\left\langle p_k\right\rangle.
	\end{align}
	Thus, we find the covariance from Eqs. \eqref{eq:firstorderclickmoments} and \eqref{eq:secondorderclickmoments},
	\begin{align}\label{eq:covariance}
	\begin{aligned}
		\overline{\Delta N_{k}\Delta N_{k'}}
		=&N\left\langle p_k\right\rangle(\delta_{k,k'}-\left\langle p_{k'}\right\rangle)
		\\&+N(N-1)\left\langle \Delta p_k\Delta p_{k'}\right\rangle.
	\end{aligned}
	\end{align}
	Note that the multinomial distribution has the covariances $\overline{\Delta N_{k}\Delta N_{k'}}=Np_k(\delta_{k,k'}-p_{k'})$ \cite{FEHP11}.
	Multiplying Eq. \eqref{eq:covariance} with $N$ and using Eq. \eqref{eq:firstorderclickmoments}, we can introduce the $(K+1)\times(K+1)$ matrix
	\begin{align}\label{eq:DefinitionMatrix}
	\begin{aligned}
		M=&\Big(N\overline{\Delta N_{k}\Delta N_{k'}}-\overline{N_k}(N\delta_{k,k'}-\overline{N_{k'}})\Big)_{k,k'=0,\ldots,K}
		\\=&N^2(N-1)\left(\left\langle \Delta p_k\Delta p_{k'}\right\rangle\right)_{k,k'=0,\ldots,K}.
	\end{aligned}
	\end{align}

	As the covariance matrix $(\langle \Delta p_k\Delta p_{k'}\rangle)_{k,k'}$ is nonnegative for any classical probability distribution $P$, we can conclude:
	We have a nonclassical light field if
	\begin{align}\label{eq:nonclassicality}
		0\nleq\left(N\overline{\Delta N_{k}\Delta N_{k'}}{-}\overline{N_k}(N\delta_{k,k'}{-}\overline{N_{k'}})\right)_{k,k'=0,\ldots,K};
	\end{align}
	i.e., the symmetric matrix $M$ in Eq. \eqref{eq:DefinitionMatrix} has at least one negative eigenvalue.
	In other words, $M\ngeq 0$ means that fluctuations of the parameters $p_k$ in $(\langle \Delta p_k\Delta p_{k'}\rangle)_{k,k'}$ are below the classical threshold of zero.
	Based on condition \eqref{eq:nonclassicality}, we will experimentally certify nonclassicality.

\begin{figure}[ht]
	\includegraphics[width=8cm]{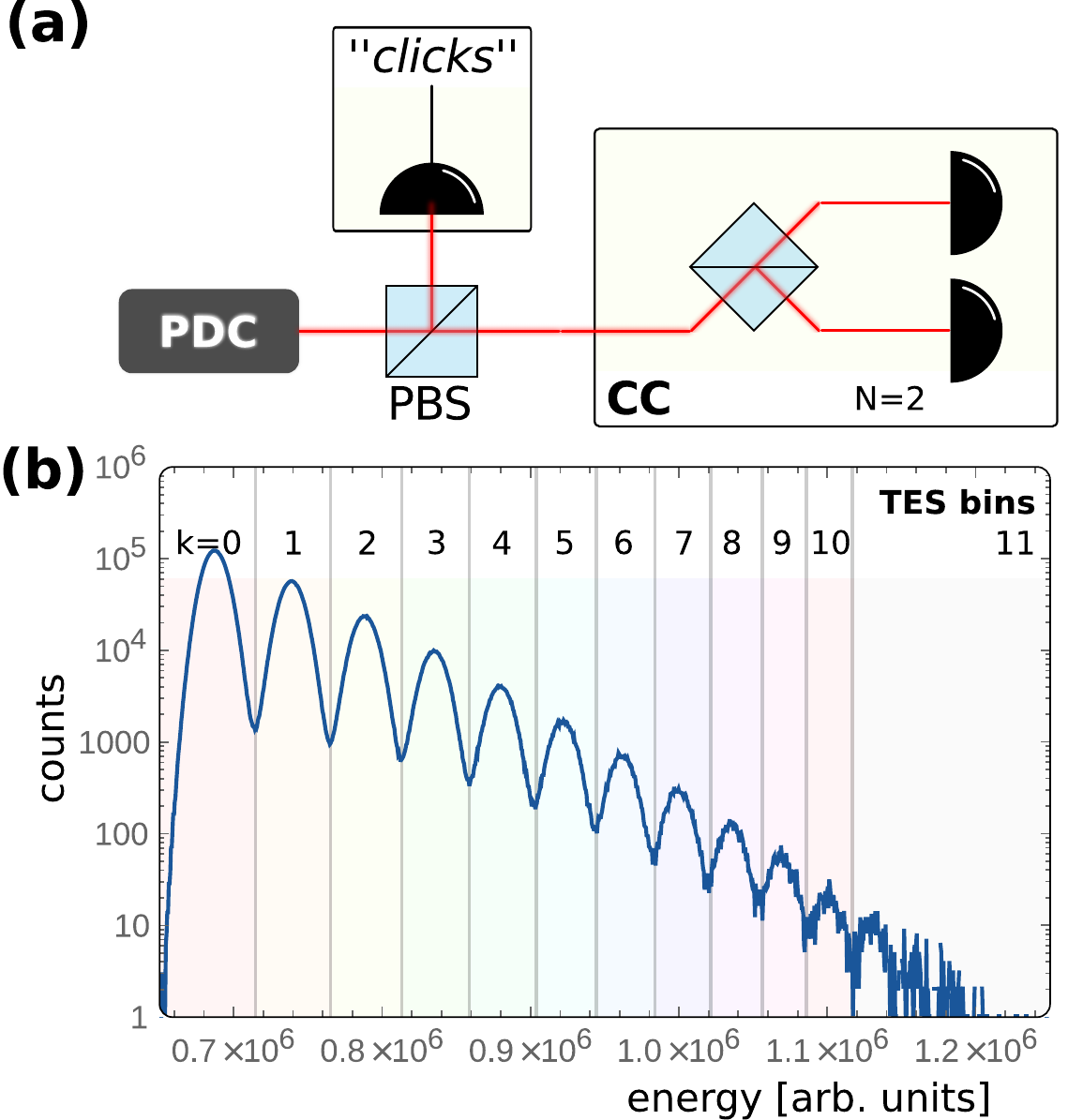}
	\caption{(Color online)
		Panel (a) depicts the experimental arrangement.
		A PDC source produces correlated photon pairs which are separated with a polarizing beam splitter (PBS).
		A conditioning to a certain outcome (labeled as ``click'') of a single TES yields a certain number of photons in the other beam.
		The latter signal is measured with a multiplexing scheme that consists of $N=2$ TESs [cf. Fig. \ref{fig:theory}].
		Panel (b) shows the binning into $K+1$ possible outcomes (bins).
		The energies that are counted with a TES (shown for the heralding detector) can be separated into $12$ bins.
	}\label{fig:setup}
\end{figure}

\paragraph*{Experimental setup.---}
	Our experimental implementation is shown in Fig. \ref{fig:setup}(a).
	A PDC source produces correlated photons.
	Conditioned on the detection of $k$ clicks from the heralding detector, we measure the click-counting statistics $c(N_0,\ldots,N_K)$, Eq. \eqref{eq:AveragedMultinomialDistribution}.
	The key components of our experiment are (i) the PDC source and (ii) the three TESs used as our heralding detector and as our two individual detectors after the multiplexing step.
	
	(i) PDC source.
	Our PDC source is a waveguide-written $8\,\mathrm{mm}$-long periodically poled potassium titanyl phosphate crystal.
	We pump a type-II spontaneous PDC process with laser pulses at $775\,\mathrm{nm}$ and a full width at half maximum of $2\,\mathrm{nm}$ at a repetition rate of $75\,\mathrm{kHz}$.
	The heralding idler mode (horizontal polarization) is centered at $1554\,\mathrm{nm}$, while the signal mode (vertical polarization) is centered at $1546\,\mathrm{nm}$.
	The output signal and idler pulses are spatially separated with a PBS.
	The pump beam is discarded using an edge filter.
	Subsequently, the other beams are filtered by $3\,\mathrm{nm}$ bandpass filters in order to filter out the broadband background which is typically generated in dielectric nonlinear waveguides \cite{ECMS11}.

	(ii) TES detectors.
	We use superconducting TESs \cite{LMN08} as our detectors.
	They consist of $25\,\mathrm{\mu m}\times25\,\mathrm{\mu m}\times 20\,\mathrm{nm}$ slabs of tungsten inside an optical cavity designed to maximize absorption at $1500\,\mathrm{nm}$.
	They are maintained at their transition temperature by Joule heating caused by a voltage bias, which is self-stabilized via an electrothermal feedback effect \cite{I95}.
	When photons are absorbed, the increase in temperature causes a corresponding electrical signal which is picked up and amplified by a superconducting quantum interference device (SQUID) module and  amplified at room temperature.
	This results in complex time-varying signals of about $5\,\mathrm{\mu s}$ duration.
	Our TESs are operated within a dilution refrigerator with a base temperature of about $70\,\mathrm{mK}$. 
	The estimated detection efficiency is $0.98^{+0.02}_{-0.08}$ \cite{HMGHLNNDKW15}.
	The electrical throughput is measured using a waveform digitizer and assigns a bin (described below) to each output pulse \cite{Getal11}.
	We process incoming signals at a speed of up to $100\,\mathrm{kHz}$.

	The time integral of the measured signal results in an energy whose counts are shown in Fig. \ref{fig:setup}(b) for the heralding TES.
	It also indicates a complex, nonlinear response of the TESs \cite{SupplementalMaterial}.
	The energies are binned into $K+1$ different intervals.
	One typically fits those counts with a number of functions or histograms to get the photon statistics via numerical reconstruction algorithms for the particular detector.
	Our bins---also the number of them---are solely determined from the measured data by simply dividing our recorded signal into disjoint energy intervals [Fig. \ref{fig:setup}(b)].
	This does not require any detector model or reconstruction algorithms.
	Above a threshold energy, no further peaks can be significantly resolved and those events are collected in the last bin.
	No measured event is discarded.
	Our heralding TES allows for a resolution of $K+1=12$ outcomes.
	Because of the splitting of the photons on the beam splitter in the multiplexing step, the data from the other two TESs yield a reduced distinction between $K+1=8$ outcomes.

\begin{figure}[ht]
	\includegraphics[height=4.8cm]{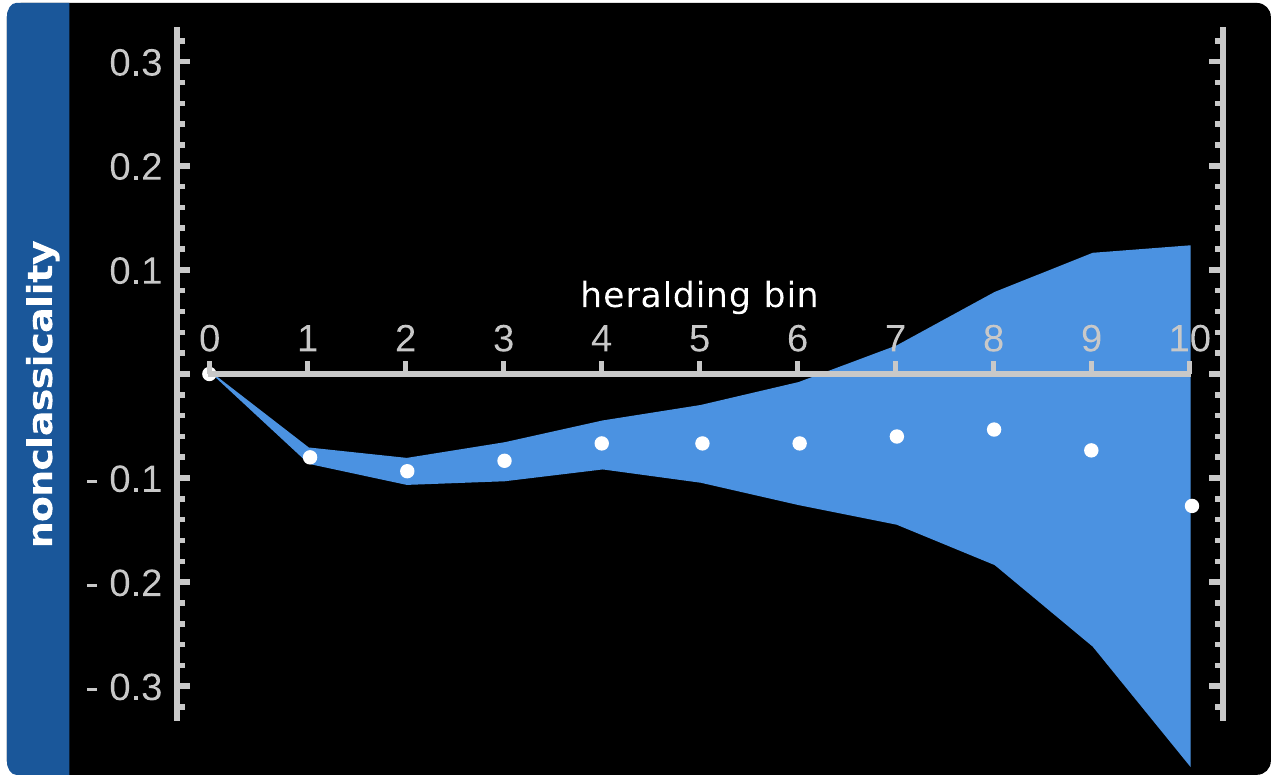}
	\caption{(Color online)
		The minimal eigenvalue of the matrix $M$ in Eq. \eqref{eq:DefinitionMatrix} is shown including its error bars (shaded area) \cite{SupplementalMaterial} as a function of the generated states, which are defined by the bin of the heralding TES.
		A negative value is inconsistent with classical optics and, therefore, verifies nonclassical light.
	}\label{fig:result}
\end{figure}

\paragraph*{Results.---}
	Condition \eqref{eq:nonclassicality} can be directly applied to the measured statistics $c(N_0,\ldots,N_K)$ by sampling mean values, variances, and covariances [Eq. \eqref{eq:DefinitionMatrix}].
	In Fig. \ref{fig:result}, we show the resulting nonclassicality of the heralded states.
	As the minimal eigenvalue of $M$ has to be non-negative for classical light, this eigenvalue is depicted in Fig. \ref{fig:result}.

	To discuss our results, we compare our findings with a simple, idealized model.
	Our produced PDC state can be approximated by a two-mode squeezed-vacuum state which has a correlated photon statistics, $p(n,n')=(1-\lambda)\lambda^{n}\delta_{n,n'}$, where $n$($n'$) is the signal(idler) photon number and $r\geq 0$ ($\lambda=\tanh^2r$) is the squeezing parameter which is a function of the pump power of the PDC process \cite{A13}.
	Heralding with an ideal PNR detector, which can resolve any photon number with a finite efficiency $\tilde\eta$, we get a conditioned statistics of the form
	\begin{align}\label{eq:model}
	\begin{aligned}
		p(n|k)=&\mathcal N_{k}\binom{n}{k} \tilde\eta^{k}(1-\tilde\eta)^{n-k}(1-\lambda)\lambda^{n},
		\\\text{with }
		\mathcal N_{k}=&\frac{(1-\lambda)(\lambda\tilde \eta)^k}{[1-\lambda(1-\tilde\eta)]^{k+1}},
	\end{aligned}
	\end{align}
	for the $k$th heralded state and $p(n|k)=0$ for $n<k$ and $\lambda^0=1$.
	Here $\mathcal N_k$ is a normalization constant as well as the probability that the $k$th state is realized.
	The signal includes at least $n\geq k$ photons if $k$ photoelectric counts have been recorded by the heralding detector.

	In the ideal case, the heralding to the 0th bin yields a thermal state [Eq. \eqref{eq:model}] and in the limit of vanishing squeezing a vacuum state, $p(n|0)=\delta_{n,0}$ for $\lambda\to 0$.
	Hence, we expect that the measured statistics is close to a multinomial, which implies $M\approx 0$.
	Our data are consistent with this consideration, cf. Fig. \ref{fig:result}.

	Using an ideal detector, a heralding to higher bin numbers would give a nonclassical Fock state with the corresponding photon number.
	The nonclassical character of the experimentally realized multiphoton states is certified in Fig. \ref{fig:result}.
	The generation of $k$ photon pairs in the PDC is less likely for higher photon numbers, $\mathcal N_k\propto \lambda^k$.
	Hence, this reduced count rate of events results in the increasing contribution of the statistical error in Fig. \ref{fig:result}.
	The highest significance of nonclassicality is found for lower heralding bins.

\begin{figure}[ht]
	\includegraphics[height=4.8cm]{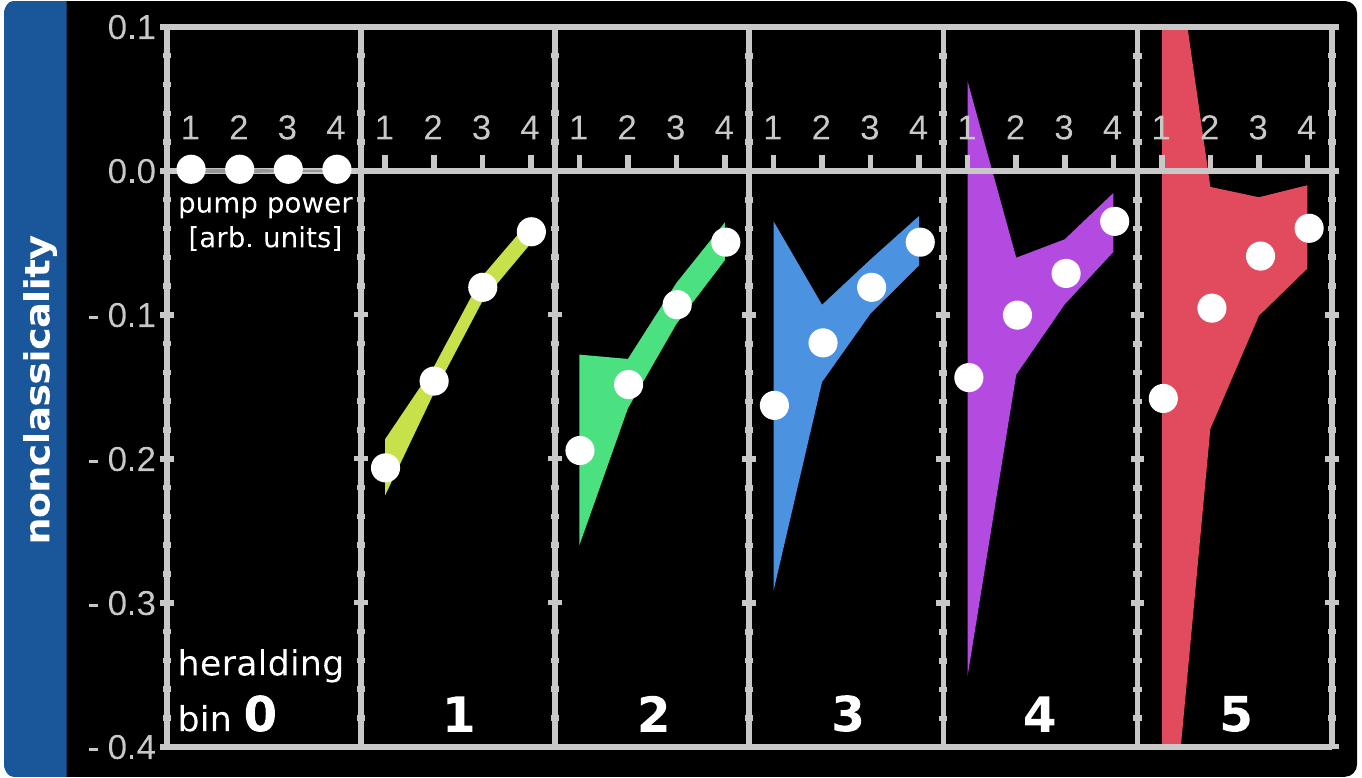}
	\caption{(Color online)
		The minimal eigenvalue of $M$ of the first six heralded states is shown as a function of the pump power.
		The nonclassicality (negative values) decreases with increasing power.
		However, the verification is more significant for higher pump powers.
	}\label{fig:result-multi}
\end{figure}

	Furthermore, we studied our criterion \eqref{eq:nonclassicality} as a function of the pump power in Fig. \ref{fig:result-multi} to demonstrate its impact on the nonclassicality.
	The conditioning to zero clicks of the heralding TES is consistent with a classical signal.
	For higher heralding bins, we observe that the nonclassicality is larger for decreasing pump powers as the distribution in Eq. \eqref{eq:model} becomes closer to a pure Fock state.
	We can also observe in Fig. \ref{fig:result-multi} that the error is larger for smaller pump powers as fewer photon pairs are generated ($\mathcal N_k\propto \lambda^k$).

	Note that the nonclassicality is expressed in terms of the photon-number correlations.
	If our detector would allow for a phase resolution, we could observe the increase of squeezing with increasing pump power.
	This suggests a future enhancement of the current setup.
	Moreover, an implementation of multiple multiplexing steps ($N>2$) would allow one to measure higher-order moments \cite{TheArticle}, which renders it possible to certify nonclassicality beyond second-order moments \cite{AT92,SBVHBAS15,APHM16}.

\paragraph*{Conclusions.---}
	We have formulated and implemented a robust and easily accessible method that can be applied to verify nonclassical light with arbitrary detectors.
	Based on a multiplexing layout, we showed that a mixture of multinomial distributions describes the measured statistics in classical optics independently of the specific properties of the individual detectors.
	We derived classical bounds on the covariance matrix whose violation is a clear signature of nonclassical light.
	We applied our theory to an experiment consisting of a single multiplexing step and two superconducting transition-edge sensors.
	We successfully demonstrated the nonclassicality of heralded multiphoton states.
	We also studied the dependence on the pump power of our spontaneous parametric-down-conversion light source.

	Our method is a straightforward technique that also applies to, e.g., temporal multiplexing or other types of individual detectors, e.g., multipixel cameras.
	It also includes the approach for avalanche photodiodes \cite{SVA12,BDJDBW13} in the special case of a binary outcome.
	Because our theory applies to general detectors, one challenge was to apply it to superconducting transition-edge sensors whose characteristics are less well understood than those of commercially available detectors.
	Our nonclassicality analysis is only based on covariances between different outcomes which requires neither sophisticated data processing nor a lot of computational time.
	Hence, it presents a simple and yet reliable tool for characterizing nonclassical light for applications in quantum technologies.

\paragraph*{Acknowledgements.---}
	The project leading to this application has received funding from the European Union's Horizon 2020 research and innovation programme under grant agreement No. 665148 (QCUMbER).
	A.~E. is supported by EPSRC EP/K034480/1.
	J.~J.~R. is supported by the Netherlands Organization for Scientific Research (NWO).
	W.~S.~K. is supported by EPSRC EP/M013243/1.
	S.~W.~N., A.~L., and T.~G. are supported by the Quantum Information Science Initiative (QISI).
	I.~A.~W. acknowledges an ERC Advanced Grant (MOQUACINO).
	The authors thank Johan Fopma for technical support.
	The authors gratefully acknowledge helpful comments by Tim Bartley and Omar Maga\~{n}a-Loaiza.

\paragraph*{Note.---}
	This work includes contributions of the National Institute of Standards and Technology, which are not subject to U.S. copyright.


\onecolumngrid
\includegraphics[width=\textwidth]{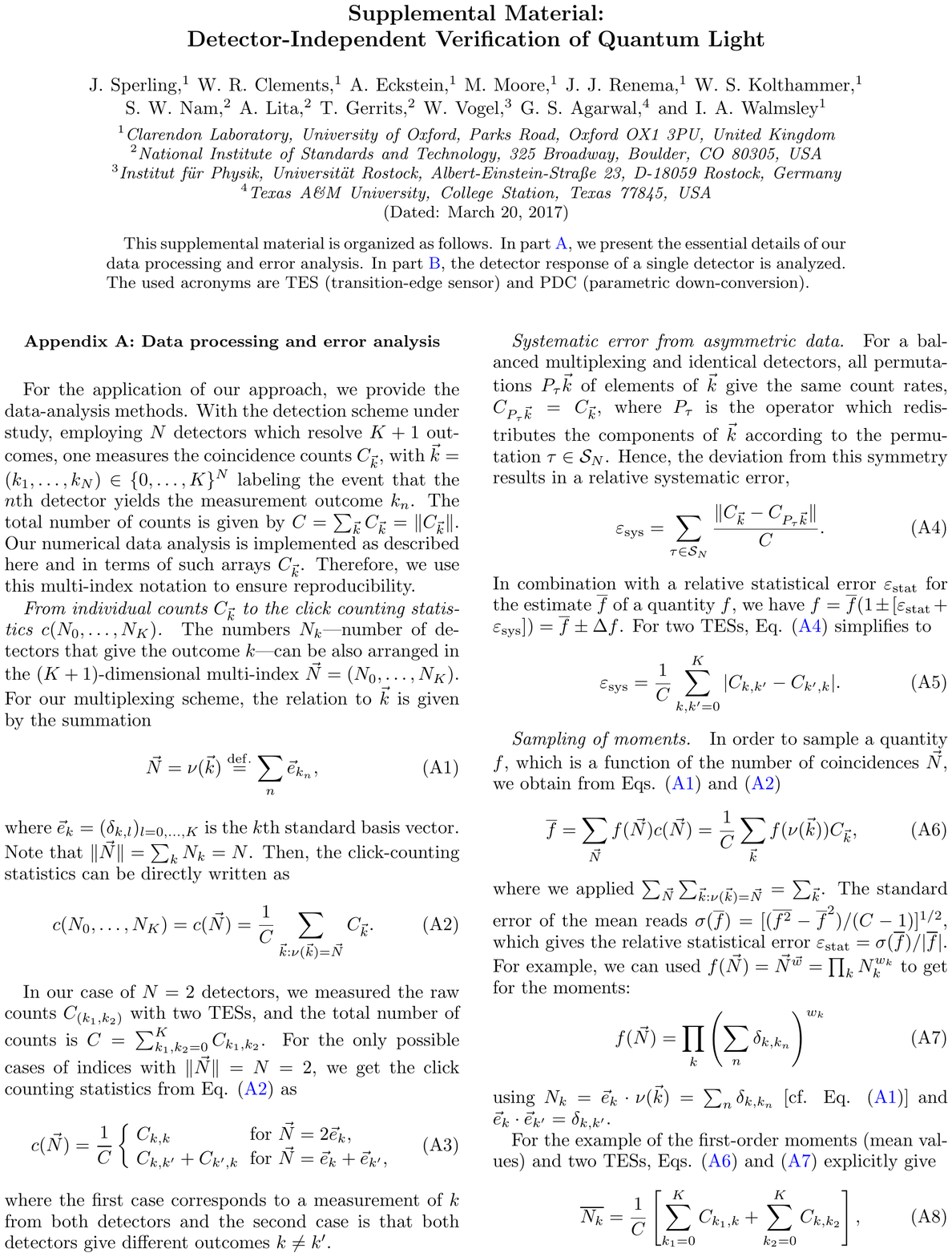}
\includegraphics[width=\textwidth]{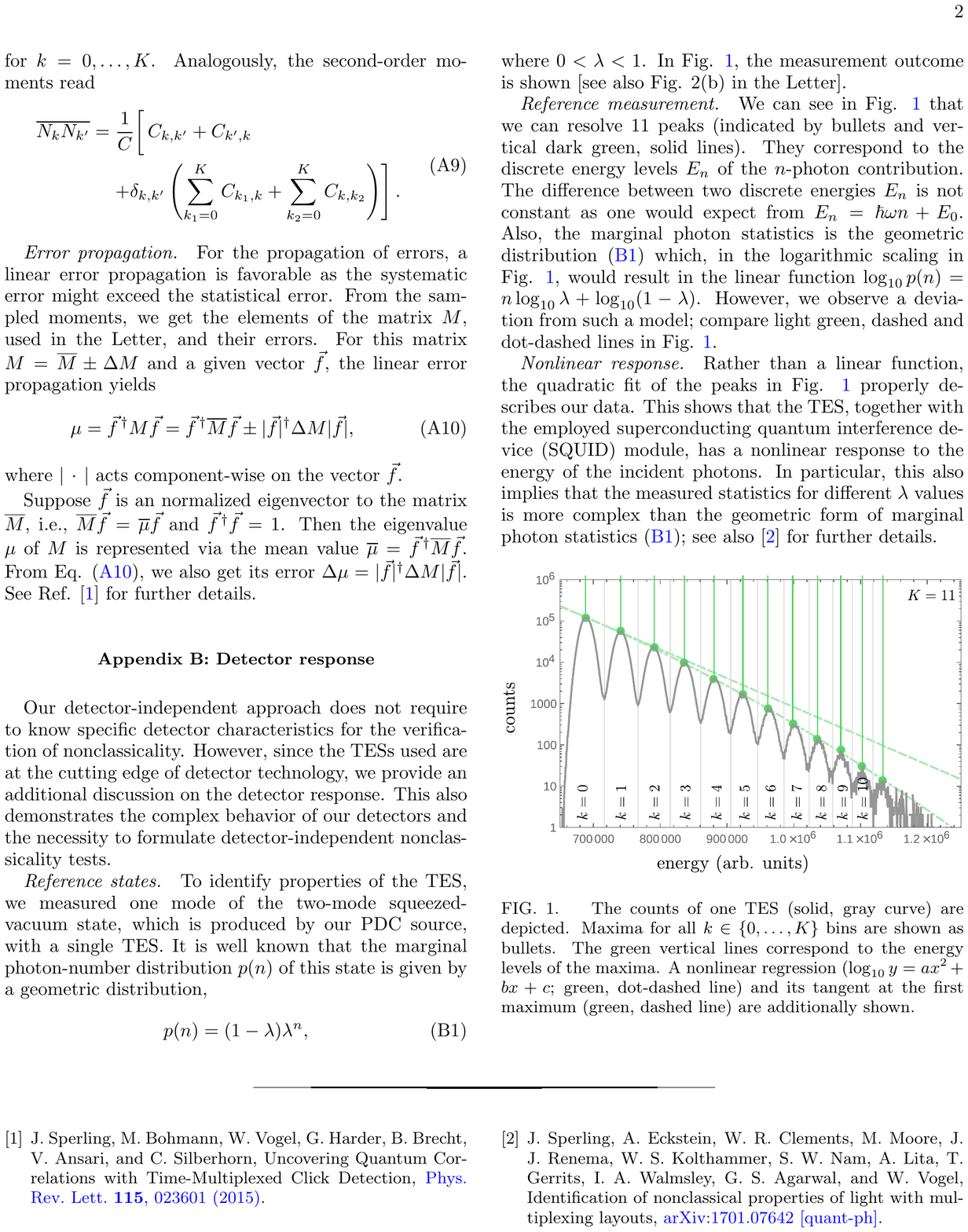}


\begin{thebibliography}{99}
\bibitem{CommentBin}
		The notion of a bin is used here synonymous with measurement outcome.
		It should not be confused with the concept of a temporal bin, which is used to describe time-bin multiplexing detectors \cite{ASSBW03,FJPF03}.
\end{thebibliography}
\end{document}